\documentstyle[multicol,eqsecnum,aps]{revtex}
\renewcommand{\narrowtext}{\begin{multicols}{2}
\global\columnwidth20.5pc\noindent}
\renewcommand{\widetext}{\end{multicols}
\global\columnwidth42.5pc}
\begin{document}
\title{Novel Density-Wave States of Two-Band Peierls-Hubbard Chains}
\author{Shoji Yamamoto}
\address{Department of Physics, Faculty of Science,
         Okayama University,\\
         Tsushima, Okayama 700-8530, Japan}
\date{Received 11 May 1998}
\maketitle
\begin{abstract}
   Based on a symmetry argument we systematically reveal
Hartree-Fock broken-symmetry solutions of the one-dimensional
two-band extended Peierls-Hubbard model, which covers various
materials of interest such as halogen-bridged metal complexes
and mixed-stack charge-transfer salts.
We find out all the regular-density-wave solutions with an ordering
vector $q=0$ or $q=\pi$.
Changing band filling as well as electron-electron and
electron-phonon interactions, we numerically inquire further into
the ground-state phase diagram and the physical property of each
state.
The possibility of novel density-wave states appearing is argued.
\end{abstract}

\narrowtext
\section{Introduction}\label{S:I}

   One-dimensional two-band models with competing electron-electron
(el-el) and electron-phonon (el-ph) interactions have been attracting
considerable interest.
Such models cover quasi-one-dimensional materials such as
halogen-bridged transition-metal (MX) linear-chain complexes and
charge-transfer (CT) salts with mixed-stacks of alternating donor
and acceptor molecules.
They are also one-dimensional analogs \cite{Gamm58} of the multiband
models for CuO$_2$ planes in high-temperature superconductors.
We retain electronic orbitals of different kinds in our calculation
not only due to a quantitative explanation of experiments but
also because we are fascinated by the models themselves which could
bring us unusual phenomena beyond the single-band description
\cite{Baer39,Nasu65}.
One of the most interesting consequences of intrinsic multiband
effects and the competition between el-el and el-ph interactions
may be the variety of ground states.
Employing a group-theoretical technique the present author and
Ozaki \cite{Yama29,Yama99} studied the three-band Peierls-Hubbard
model on a square lattice and obtained plenty of novel broken
symmetry phases.
Gammel {\it et al.} \cite{Gamm08} investigated a similar model but
with both site-diagonal and site-off-diagonal el-ph interactions.
Besides regular-density-wave states \cite{Yone65}, they pointed out
several novel solutions such as incommensurate long-period phases
\cite{Bish51} and a spin-frustrated state \cite{Gamm77}.

   Some of the ground states predicted above were observed
experimentally.
Charge-transfer compounds with mixed stacks such as
TTF-chloranil \cite{Torr53} and TTeC$_1$TTF-TCNQ \cite{Iwas15}
may be described by a half-filled two-band model with
site-off-diagonal el-ph coupling.
They are dimerized into bond-order-wave (BOW) states at low
temperatures, which is recognized as a charge-transfer-induced
spin-Peierls transition.
Halogen-bridged metal complexes, which may be described by a
$3/4$-filled two-band model with site-diagonal el-ph coupling,
exhibit a strong dependence of their ground states on the constituent
metal and halogen ions.
Familiar compounds of Pt ions \cite{Kell57} have the ground state
with a charge density wave (CDW) on the metal sites induced by a
dimerization of the halogen-ion sublattice, whereas recently
synthesized Ni compounds \cite{Toft61,Tori41} show a regular-chain
structure with a spin density wave (SDW) on the metal sites.
Although the interchain hydrogen-bond networks are not negligible
here, the competition between the CDW and SDW states may be
recognized as a crossover between the Peierls and Mott insulators.
Thus observed rich phase diagrams allow us to expect further ground
states unrevealed.
Hence we here present a systematic Hartree-Fock (HF) study on
competing broken-symmetry ground states.
We focus on one dimension where ground-state properties could be
most fascinating due to significant quantum effects.
Based on a symmetry argument, we reveal possible ground states of
regular-density-wave type and clarify their physical properties.

\section{Model Hamiltonian and Its Symmetry Property}\label{S:H}

   We study the one-dimensional two-band extended Peierls-Hubbard
model:
\widetext
\begin{eqnarray}
   {\cal H}
   &=& \sum_{n=1}^L\sum_{s=\pm}
       \left[\varepsilon_{\rm M}-\beta(u_n-u_{n-1})\right]
       a_{1:n,s}^\dagger a_{1:n,s}
      +\sum_{n=1}^L\sum_{s=\pm}
       \varepsilon_{\rm X} a_{2:n,s}^\dagger a_{2:n,s}
       \nonumber \\
   &-& \sum_{n=1}^L\sum_{s=\pm}
       \left[(t-\alpha u_n)a_{1:n,s}^\dagger a_{2:n,s}
       +(t+\alpha u_{n-1})a_{1:n,s}^\dagger a_{2:n-1,s}
       +{\rm H.c.}\right]
       \nonumber \\
   &+& U_{\rm M}\sum_{n=1}^L
       a_{1:n,+}^\dagger a_{1:n,+}
       a_{1:n,-}^\dagger a_{1:n,-}
      +U_{\rm X}\sum_{n=1}^L
       a_{2:n,+}^\dagger a_{2:n,+}
       a_{2:n,-}^\dagger a_{2:n,-}
       \nonumber \\
   &+& V\sum_{n=1}^L\sum_{s,s'=\pm}
       \left[
        a_{1:n,s}^\dagger a_{1:n,s} a_{2:n,s'}^\dagger a_{2:n,s'}
       +a_{1:n,s}^\dagger a_{1:n,s} a_{2:n-1,s'}^\dagger a_{2:n-1,s'}
       \right]
    +  \frac{K}{2}\sum_{n=1}^L
       u_n^2 \,.
   \label{E:Hn}
\end{eqnarray}
\narrowtext
Here we assume the Hamiltonian to describe MX chains.
Now the Hamiltonian (\ref{E:Hn}) is recognized as follows:
$a_{1:n,s}^\dagger$ and $a_{2:n,s}^\dagger$ are the creation
operators of an electron with spin $s=\pm$ (up and down respectively)
in the metal $d_{z^2}$ and halogen $p_z$ orbitals at unit cell
$n$, respectively, $u_n$ the chain-direction displacement from the
uniform lattice spacing of the halogen atom at unit cell $n$, and $L$
the number of unit cells composed of a pair of neighboring metal
and halogen atoms;
$\varepsilon_{\rm M}$ and $\varepsilon_{\rm X}$
are the energies of the metal $d_{z^2}$ and halogen $p_z$
orbitals on isolated atoms, respectively,
$t$ the transfer energy of hopping between these levels,
$U_{\rm M}$, $U_{\rm X}$, and $V$
the el-el interactions on the metal atoms, on the halogen atoms, and
between the neighboring metal and halogen atoms, respectively;
$\alpha$, $\beta$, and $K$ denote
the site-off-diagonal and site-diagonal el-ph coupling constants
and the elastic constant, respectively.
We stress that the model essentially covers various materials.
For example, replacing the $d_{z^2}$ and $p_z$ orbitals by $p\pi$
orbitals of donor and acceptor molecules, respectively, we can
immediately switch our argument to organic mixed-stack CT compounds.
Fourier transforms
\begin{eqnarray}
   a_{1:n,s}^\dagger
      &=&\frac{1}{\sqrt{L}}\sum_k{\rm e}^{-{\rm i}kn}a_{1:k,s}\,,
   \nonumber \\
   a_{2:n,s}^\dagger
      &=&\frac{1}{\sqrt{L}}\sum_k{\rm e}^{-{\rm i}k(n+1/2)}a_{2:k,s}\,,
   \nonumber \\
   u_n
      &=&\frac{1}{\sqrt{L}}\sum_k{\rm e}^{-{\rm i}k(n+1/2)}u_k\,,
   \label{E:Four}
\end{eqnarray}
compactly rewrite the Hamiltonian as
\widetext
\begin{eqnarray}
   {\cal H}
   &=& \sum_{i,j}\sum_{k,q}\sum_{s,s'}
       \langle i:k+q,s\vert t\vert j:k,s'\rangle
       a_{i:k+q,s}^\dagger a_{j:k,s'}
       \nonumber \\
   &+& \frac{1}{2}\sum_{i,j,m,n}\sum_{k,k',q}\sum_{s,s',t,t'}
       \langle i:k+q,s;m:k',t\vert v\vert j:k,s';n:k'+q,t'\rangle
       \nonumber \\
   & & \times
       a_{i:k+q,s}^\dagger a_{m:k',t}^\dagger
       a_{n:k'+q,t'} a_{j:k,s'}
    +  \frac{K}{2}
       \sum_k u_k u_k^* \,,
   \label{E:Hk}
\end{eqnarray}
\narrowtext
where
\begin{eqnarray}
   && \langle i:k+q,s\vert t\vert j:k,s'\rangle
     =\langle i:k+q\vert t\vert j:k\rangle\delta_{ss'} \,,
      \label{E:intt}
      \\
   && \langle i:k+q,s;m:k',t\vert v\vert j:k,s';n:k'+q,t'\rangle
      \nonumber \\
   && \quad
     =\langle i:k+q;m:k'\vert v\vert j:k;n:k'+q\rangle
      \delta_{ss'}\delta_{tt'} \,,
      \label{E:intv}
\end{eqnarray}
are specified as
\begin{eqnarray}
   && \langle 1:k+q\vert t\vert 1:k\rangle
     ={\widetilde\varepsilon}_{\rm M}\delta_{q0}
     -\frac{2{\rm i}\beta}{\sqrt{L}}u_q^*
      \sin\left(\frac{q}{2}\right) \,,
      \nonumber \\
   && \langle 2:k+q\vert t\vert 2:k\rangle
     ={\widetilde\varepsilon}_{\rm X}\delta_{q0} \,,
      \nonumber \\
   && \langle 1:k+q\vert t\vert 2:k\rangle
     =-2t_0\cos\left(\frac{k}{2}\right)\,\delta_{q0}
      \nonumber \\
   && \qquad\qquad\qquad\qquad
      +\frac{2{\rm i}\alpha}{\sqrt{L}}u_q^*
      \sin\left(\frac{k+q}{2}\right) \,,
      \\
   && \langle 1:k+q;1:k'\vert v\vert 1:k;1:k'+q\rangle
     =\frac{U_{\rm M}}{L} \,,
      \nonumber \\
   && \langle 2:k+q;2:k'\vert v\vert 2:k;2:k'+q\rangle
     =\frac{U_{\rm X}}{L} \,,
      \nonumber \\
   && \langle 1:k+q;2:k'\vert v\vert 1:k;2:k'+q\rangle
     =\frac{2V}{L}\cos\left(\frac{q}{2}\right) \,.
   \label{E:int}
\end{eqnarray}
with the renormalized on-site affinities
${\widetilde\varepsilon}_{\rm M}=\varepsilon_{\rm M}-U_{\rm M}/2$
and
${\widetilde\varepsilon}_{\rm X}=\varepsilon_{\rm X}-U_{\rm X}/2$.

   The symmetry group of the system is generally represented as
\begin{equation}
   {\bf G}={\bf P}\times{\bf S}\times{\bf T} \,,
   \label{E:G}
\end{equation}
where ${\bf P}={\bf L}_1\land{\bf C}_2$ is the space group of a linear
chain with the one-dimensional translation group ${\bf L}_1$ whose
basis vector is the unit-cell translation $l_1$,
${\bf S}$ the group of spin-rotation, and
${\bf T}$ the group of time reversal.
We here take no account of the gauge group because we do not consider
superconducting phases.
Group actions on the creation operators are
defined as follows \cite{Ozak55,Yama49}:
\begin{eqnarray}
   &&l\cdot a_{i:k,s}^\dagger={\rm e}^{-{\rm i}kl}a_{i:k,s}^\dagger\,,
   \label{E:GAl} \\
   &&p\cdot a_{i:k,s}^\dagger=a_{i:pk,s}^\dagger \,,
   \label{E:GAp} \\
   &&u(\mbox{\boldmath$e$},\theta)\cdot a_{i:k,s}^\dagger
      =\sum_{s'}
         [u(\mbox{\boldmath$e$},\theta)]_{ss'}   a_{k,s'}^\dagger\,,
   \label{E:GAu} \\
   &&t\cdot(f a_{i:k,s}^\dagger)
      =-s\,f^* a_{i:-k,-s}^\dagger\,,
   \label{E:GAt}
\end{eqnarray}
where
$l\in{\bf L}_1$,
$p\in{\bf C}_2$,
$u(\mbox{\boldmath$e$},\theta)\in{\bf S}$,
$t\in{\bf T}$, and
$f$ is an arbitrary complex number.
$
   u(\mbox{\boldmath$e$},\theta)
   =\sigma^0\cos(\theta/2)
   -(\mbox{\boldmath$\sigma$}\cdot\mbox{\boldmath$e$})
    \sin(\theta/2)
$
represents the spin rotation of angle $\theta$ around an axis
$\mbox{\boldmath$e$}$, where
$\sigma^0$ is the $2\times 2$ unit matrix and
$\mbox{\boldmath$\sigma$}=(\sigma^x,\sigma^y,\sigma^z)$ is a vector
composed of the Pauli-matrices.

   Let $\check{G}$ denote the irreducible representations of {\bf G}
over the real number field, where their representation space is spanned
by the Hermitian operators \{$a_{i:k,s}^\dagger a_{j:k',s'}$\}.
There is a one-to-one correspondence \cite{Ozak83,Ozak14} between
$\check{G}$ and broken-symmetry phases with a definite ordering vector.
Any representation $\check{G}$ is obtained as a Kronecker product of
the irreducible representations of ${\bf P}$, ${\bf S}$, and ${\bf T}$
over the real number field:
\begin{equation}
   \check{G}=\check{P}\otimes\check{S}\otimes\check{T} \,.
   \label{E:Grep}
\end{equation}
$\check{P}$ is characterized by an ordering vector $q$ in the
Brillouin zone and an irreducible representation of its little group
${\bf P}(q)$ \cite{Brad}, and is therefore labeled $q\check{P}(q)$.
The relevant representations of ${\bf S}$ and ${\bf T}$ are,
respectively, given by
\begin{eqnarray}
   &&
   \check{S}^0(u(\mbox{\boldmath$e$},\theta))
      =1 \,,\ \ 
   \check{S}^1(u(\mbox{\boldmath$e$},\theta))
      =O(u(\mbox{\boldmath$e$},\theta)) \,,
   \label{E:S} \\
   &&
   \check{T}^0(t)=1\,,\ \ 
   \check{T}^1(t)=-1\,,
   \label{E:T}
\end{eqnarray}
where $O(u(\mbox{\boldmath$e$},\theta))$ is the $3\times 3$
orthogonal matrix satisfying
$
   u(\mbox{\boldmath$e$},\theta)
   \mbox{\boldmath$\sigma$}^\lambda
   u^\dagger(\mbox{\boldmath$e$},\theta)
      =\sum_{\mu=x,y,z}
       [O(u(\mbox{\boldmath$e$},\theta))]_{\lambda\mu}
       \mbox{\boldmath$\sigma$}^\mu \ \
       (\lambda=x,\,y,\,z)
$.
The representations
$\check{P}\otimes\check{S}^0\otimes\check{T}^0$,
$\check{P}\otimes\check{S}^1\otimes\check{T}^1$,
$\check{P}\otimes\check{S}^0\otimes\check{T}^1$, and 
$\check{P}\otimes\check{S}^1\otimes\check{T}^0$
correspond to charge-density-wave (CDW), spin-density-wave (SDW),
charge-current-wave (CCW), and spin-current-wave (SCW) states,
respectively.
We leave out current-wave states in our argument, because here in
one dimension all of them but one-way uniform-current states break
the charge- or spin-conservation law.
The current states in one dimension are less interesting than in
higher dimensions, where physically stimulative ones
\cite{Yama29,Ozak55,Yama49,Affl74} may appear.
We consider two ordering vectors $q=0$ and $q=\pi$, which are labeled
$\mit\Gamma$ and $X$, respectively.
Thus we treat the instabilities characterized as
${\mit\Gamma}\check{P}({\mit\Gamma})
 \otimes\check{S}^0\otimes\check{T}^0$,
$X\check{P}(X)
 \otimes\check{S}^0\otimes\check{T}^0$,
${\mit\Gamma}\check{P}({\mit\Gamma})
 \otimes\check{S}^1\otimes\check{T}^1$, and
$X\check{P}(X)
 \otimes\check{S}^1\otimes\check{T}^1$.
Here
$\check{P}({\mit\Gamma})$ and $\check{P}(X)$ are either $A$
($C_2$-symmetric) or $B$ ($C_2$-antisymmetric) representation of
${\bf C}_2$ because ${\bf P}({\mit\Gamma})={\bf P}(X)={\bf C}_2$
in the present system.

\section{Broken Symmetry Solutions}\label{S:BSS}

   In the HF approximation the Hamiltonian (\ref{E:Hk}) is replaced
by
\widetext
\begin{equation}
   {\cal H}_{\rm HF}
      =\sum_{i,j}\sum_{k,s,s'}\sum_{\lambda=0,z}
       \left[
         x_{ij}^{\lambda}({\mit\Gamma};k)
         a_{i:k,s}^\dagger a_{j:k,s'}
        +x_{ij}^{\lambda}(X;k)
         a_{i:k+\pi,s}^\dagger a_{j:k,s'}
       \right]
       \sigma_{ss'}^\lambda\,,
   \label{E:HHF}
\end{equation}
where
$x_{ij}^\lambda(K;k)$ is the self-consistent renormalized field
describing nonmagnetic ($\lambda=0$) and magnetic ($\lambda=z$)
instabilities at point $K$ and is expressed as
\begin{eqnarray}
   &&
   x_{ij}^0({\mit\Gamma};k)
     =\langle i:k\vert t\vert j:k\rangle
     +\sum_{m,n}\sum_{k'}
      \rho_{nm}^0({\mit\Gamma};k')
   \nonumber \\
   &&
   \quad\times
   \left[
    2\langle i:k;m:k'\vert v\vert j:k;n:k'\rangle
    -\langle i:k;m:k'\vert v\vert n:k';j:k\rangle
   \right]\,,
   \nonumber \\
   &&
   x_{ij}^z({\mit\Gamma};k)
     =-\sum_{m,n}\sum_{k'}
      \rho_{nm}^z({\mit\Gamma};k')
      \langle i:k;m:k'\vert v\vert n:k';j:k\rangle
   \nonumber \\
   &&
   x_{ij}^0(X;k)
     =\langle i:k+\pi\vert t\vert j:k\rangle
     +\sum_{m,n}\sum_{k'}
      (-1)^{\delta_{m2}}\rho_{nm}^0({\mit\Gamma};k')
   \nonumber \\
   &&
   \quad\times
   \left[
    2\langle i:k+\pi;m:k'\vert v\vert j:k;n:k'+\pi\rangle
    -\langle i:k+\pi;m:k'\vert v\vert n:k'+\pi;j:k\rangle
   \right]\,,
   \nonumber \\
   &&
   x_{ij}^z(X;k)
   =-\sum_{m,n}\sum_{k'}
          (-1)^{\delta_{m2}}
          \rho_{nm}^z(X;k')
          \langle i:k+\pi;m:k'\vert v\vert n:k'+\pi;j:k\rangle\,.
   \label{E:SCF}
\end{eqnarray}
\narrowtext
Here we have introduced the density matrices
\begin{eqnarray}
   \rho_{ij}^\lambda({\mit\Gamma};k)
      &=& \frac{1}{2}\sum_{s,s'}
          \langle a_{j:k,s}^\dagger a_{i:k,s'}\rangle_{\rm HF}
          \sigma_{ss'}^\lambda \,,
   \nonumber \\
   \rho_{ij}^\lambda(X;k)
      &=& \frac{1}{2}\sum_{s,s'}
          \langle a_{j:k+\pi,s}^\dagger a_{i:k,s'}\rangle_{\rm HF}
          \sigma_{ss'}^\lambda \,,
   \label{E:rho}
\end{eqnarray}
where $\langle\cdots\rangle_{\rm HF}$ denotes the quantum average in
a HF eigenstate.
We note that nonaxial magnetic instabilities ($\lambda=x,y$) are not
obtained from the Hamiltonian (\ref{E:HHF}) because all the
irreducible representations of ${\bf C}_2$ are of one dimension.
${\cal H}_{\rm HF}$ is decomposed into spatial-symmetry-definite
components as
\begin{equation}
   {\cal H}_{\rm HF}
      =\sum_{K={\mit\Gamma},X}\sum_{D=A,B}\sum_{\lambda=0,z}
       h^\lambda(K;D) \,,
   \label{E:Decom}
\end{equation}
where $h^\lambda(K;D)$ is the irreducible basis Hamiltonian which
belongs to the $D$ representation at point $K$ and is specified as
\widetext
\begin{eqnarray}
   h^0({\mit\Gamma};A)
      &=& \sum_{k,s}
          c_1^0({\mit\Gamma};A)
          a_{1:k,s}^\dagger a_{1:k,s}
       +  \sum_{k,s}
          c_2^0({\mit\Gamma};A)
          a_{2:k,s}^\dagger a_{2:k,s}
          \nonumber \\
      &+& \sum_{k,s}
          \left[
          c_3^0({\mit\Gamma};A)
          \cos\left(\frac{k}{2}\right)\,a_{1:k,s}^\dagger a_{2:k,s}
         +{\rm H.c.}
          \right] \,,
          \label{E:H0GA} \\
   h^0({\mit\Gamma};B)
      &=& \sum_{k,s}
          \left[
          c_1^0({\mit\Gamma};B)
          \sin\left(\frac{k}{2}\right)\,a_{1:k,s}^\dagger a_{2:k,s}
         +{\rm H.c.}
          \right] \,,
          \label{E:H0GB} \\
   h^0(X;A)
      &=& \sum_{k,s}
          c_1^0(X;A)
          a_{1:k+\pi,s}^\dagger a_{1:k,s}
       +  \sum_{k,s}
          \left[
          c_2^0(X;A)
          \cos\left(\frac{k}{2}\right)\,a_{1:k+\pi,s}^\dagger a_{2:k,s}
         +{\rm H.c.}
          \right] \,,
          \label{E:H0XA}
          \\
   h^0(X;B)
      &=& \sum_{k,s}
          c_1^0(X;B)
          a_{2:k+\pi,s}^\dagger a_{2:k,s}
       +  \sum_{k,s}
          \left[
          c_2^0(X;B)
          \sin\left(\frac{k}{2}\right)\,a_{1:k+\pi,s}^\dagger a_{2:k,s}
         +{\rm H.c.}
          \right] \,,
          \label{E:H0XB} \\
   h^z({\mit\Gamma};A)
      &=& \sum_{k,s,s'}
          c_1^z({\mit\Gamma};A)
          a_{1:k,s}^\dagger a_{1:k,s'}\sigma_{ss'}^z
       +  \sum_{k,s,s'}
          c_2^z({\mit\Gamma};A)
          a_{2:k,s}^\dagger a_{2:k,s'}\sigma_{ss'}^z
          \nonumber \\
      &+& \sum_{k,s,s'}
          \left[
          c_3^z({\mit\Gamma};A)
          \cos\left(\frac{k}{2}\right)\,
          a_{1:k,s}^\dagger a_{2:k,s'}\sigma_{ss'}^z
         +{\rm H.c.}
          \right] \,,
          \label{E:HzGA} \\
   h^z({\mit\Gamma};B)
      &=& \sum_{k,s,s'}
          \left[
          c_1^z({\mit\Gamma};B)
          \sin\left(\frac{k}{2}\right)\,
          a_{1:k,s}^\dagger a_{2:k,s'}\sigma_{ss'}^z
         +{\rm H.c.}
          \right] \,,
          \label{E:HzGB} \\
   h^z(X;A)
      &=& \sum_{k,s,s'}
          c_1^z(X;A)
          a_{1:k+\pi,s}^\dagger a_{1:k,s'}\sigma_{ss'}^z
       +  \sum_{k,s,s'}
          \left[
          c_2^z(X;A)
          \cos\left(\frac{k}{2}\right)\,
          a_{1:k+\pi,s}^\dagger a_{2:k,s'}\sigma_{ss'}^z
         +{\rm H.c.}
          \right] \,,
          \label{E:HzXA} \\
   h^z(X;B)
      &=& \sum_{k,s,s'}
          c_1^z(X;B)
          a_{2:k+\pi,s}^\dagger a_{2:k,s'}
       +  \sum_{k,s,s'}
          \left[
          c_2^z(X;B)
          \sin\left(\frac{k}{2}\right)\,
          a_{1:k+\pi,s}^\dagger a_{2:k,s'}\sigma_{ss'}^z
         +{\rm H.c.}
          \right] \,,
          \label{E:HzXB}
\end{eqnarray}
\narrowtext
with
\begin{eqnarray}
   &&
   c_1^0({\mit\Gamma};A)
       =  \varepsilon_{\rm M}
         +\frac{U_{\rm M}}{L}\sum_{k}\rho_{11}^0({\mit\Gamma};k)
   \nonumber \\
   && \qquad\qquad\qquad\ 
         +\frac{4V}{L}\sum_{k}\rho_{22}^0({\mit\Gamma};k) \,,
   \nonumber \\
   &&
   c_2^0({\mit\Gamma};A)
       =  \varepsilon_{\rm X}
         +\frac{U_{\rm X}}{L}\sum_{k}\rho_{22}^0({\mit\Gamma};k)
   \nonumber \\
   && \qquad\qquad\qquad\ 
         +\frac{4V}{L} \sum_{k}\rho_{11}^0({\mit\Gamma};k) \,,
   \nonumber \\
   &&
   c_3^0({\mit\Gamma};A)
       = -2t_0
         -\frac{2V}{L}\sum_{k}\cos\left(\frac{k}{2}\right)\,
          \rho_{12}^0({\mit\Gamma};k) \,,
   \label{E:c0GA} \\
   &&
   c_1^0({\mit\Gamma};B)
       =  \frac{2{\rm i}\alpha}{\sqrt{L}}u_0
         -\frac{2V}{L}\sum_{k}\sin\left(\frac{k}{2}\right)\,
          \rho_{12}^0({\mit\Gamma};k) \,,
   \label{E:c0GB} \\
   &&
   c_1^0(X;A)
       =  \frac{2{\rm i}\beta}{\sqrt{L}}u_\pi
         +\frac{U_{\rm M}}{L}\sum_{k}\rho_{11}^0({\mit\Gamma};k) \,,
   \nonumber \\
   &&
   c_2^0(X;A)
       = -\frac{2{\rm i}\alpha}{\sqrt{L}}u_\pi
   \nonumber \\
   && \qquad\qquad
         +\frac{2V}{L}\sum_{k}\cos\left(\frac{k}{2}\right)\,
          \rho_{12}^0(X;k+\pi) \,,
   \label{E:c0XA} \\
   &&
   c_1^0(X;B)
       = -\frac{U_{\rm X}}{L}\sum_{k}\rho_{22}^0(X;k+\pi) \,,
   \nonumber \\
   &&
   c_2^0(X;B)
       =  \frac{2V}{L}\sum_{k}\sin\left(\frac{k}{2}\right)\,
          \rho_{12}^0(X;k+\pi) \,,
   \label{E:c0XB} \\
   &&
   c_1^z({\mit\Gamma};A)
       = -\frac{U_{\rm M}}{L}\sum_{k}\rho_{11}^z({\mit\Gamma};k) \,,
   \nonumber \\
   &&
   c_2^z({\mit\Gamma};A)
       = -\frac{U_{\rm X}}{L}\sum_{k}\rho_{22}^z({\mit\Gamma};k) \,,
   \nonumber \\
   &&
   c_3^z({\mit\Gamma};A)
       = -\frac{2V}{L}\sum_{k}\cos\left(\frac{k}{2}\right)\,
          \rho_{12}^z({\mit\Gamma};k) \,,
   \label{E:czGA} \\
   &&
   c_1^z({\mit\Gamma};B)
       = -\frac{2V}{L}\sum_{k}\sin\left(\frac{k}{2}\right)\,
          \rho_{12}^z({\mit\Gamma};k) \,,
   \label{E:czGB} \\
   &&
   c_1^z(X;A)
       = -\frac{U_{\rm M}}{L}\sum_{k}\rho_{11}^z(X;k+\pi) \,,
   \nonumber \\
   &&
   c_2^z(X;A)
       =  \frac{2V}{L}\sum_{k}\cos\left(\frac{k}{2}\right)\,
          \rho_{12}^z(X;k+\pi) \,,
   \label{E:czXA} \\
   &&
   c_1^z(X;B)
       =  \frac{U_{\rm X}}{L}\sum_{k}\rho_{22}^z(X;k+\pi) \,,
   \nonumber \\
   &&
   c_2^z(X;B)
       =  \frac{2V}{L}\sum_{k}\sin\left(\frac{k}{2}\right)\,
          \rho_{12}^z(X;k+\pi) \,.
   \label{E:czXB}
\end{eqnarray}
We list in Table \ref{T:H} the thus-obtained HF Hamiltonians of
broken symmetry, where their invariance groups are also displayed.
Once a broken-symmetry Hamiltonian is given, its invariance group
is defined in such a way that any symmetry operation in the group
keeps the Hamiltonian unchanged.
Keeping in mind that the density matrices characteristic of the
Hamiltonian given should also be invariant for its invariance group,
we can completely determine qualitative properties of the solution
\cite{Ozak55,Yama49}.

   Let us introduce relevant order parameters.
The local charge density on site $i$ at the $n$th unit cell is
defined as
\begin{equation}
   d_{i:n}
     \equiv
     \sum_s\langle a_{i:n,s}^\dagger a_{i:n,s}\rangle_{\rm HF}\,,
   \label{E:OPd}
\end{equation}
the local spin density on site $i$ at the $n$th unit cell as
\begin{equation}
   s_{i:n}^z
     \equiv
     \frac{1}{2}
     \sum_{s,s'}
     \langle a_{i:n,s}^\dagger a_{i:n,s'}\rangle_{\rm HF}
     \sigma_{ss'}^z\,,
   \label{E:OPs}
\end{equation}
the complex bond order between site $i$ at the $n$th unit cell
and site $j$ at the $m$th unit cell as
\begin{equation}
   p_{i:n;j:m}
     \equiv
     \sum_s
     \langle a_{i:n,s}^\dagger a_{j:m,s}\rangle_{\rm HF}\,,
   \label{E:OPp}
\end{equation}
and the complex spin bond order between site $i$ at the $n$th
unit cell and site $j$ at the $m$th unit cell as
\begin{equation}
   t_{i:n;j:m}^z
     \equiv
     \frac{1}{2}
     \sum_{s,s'}
     \langle a_{i:n,s}^\dagger a_{j:n,s'}\rangle_{\rm HF}
     \sigma_{ss'}^z\,.
   \label{E:OPt}
\end{equation}
The halogen-atom displacements $u_n$ are self-consistently determined
so as to minimize the HF energy 
$E_{\rm HF}\equiv\langle{\cal H}\rangle_{\rm HF}$.
Now we are ready to classify and characterize all the possible
solutions.
We are convinced in the following that any order parameter
unspecified is zero and $\bar{\rho}^\lambda_{ij}(K;k)$ and
$\tilde{\rho}^\lambda_{ij}(K;k)$ denote the real and imaginary parts
of ${\rho}^\lambda_{ij}(K;k)$, respectively.
The solutions obtained are schematically shown in Fig. \ref{F:BSP}.

\widetext
\subsection{${\mit\Gamma}A\otimes\check{S}^0\otimes\check{T}^0$}

\begin{eqnarray}
   &&d_{1:n}=\frac{2}{L}\sum_k\bar{\rho}_{11}^0({\mit\Gamma};k) \,,\ \
     d_{2:n}=\frac{2}{L}\sum_k\bar{\rho}_{22}^0({\mit\Gamma};k) \,,
     \nonumber \\
   &&p_{1:n;2:n}=p_{1:n;2:n-1}
        =\frac{2}{L}\sum_k\cos\left(\frac{k}{2}\right)\,
         \bar{\rho}_{21}^0({\mit\Gamma};k)\,.
   \label{E:GA00}
\end{eqnarray}
This phase (Fig.$\!$ \ref{F:BSP}a) possesses the full symmetry
(\ref{E:G}) and is recognized as the paramagnetic state.
We abbreviate the state as PM.

\subsection{${\mit\Gamma}B\otimes\check{S}^0\otimes\check{T}^0$}

\begin{eqnarray}
   &&d_{1:n}=\frac{2}{L}\sum_k\bar{\rho}_{11}^0({\mit\Gamma};k) \,,\ \
     d_{2:n}=\frac{2}{L}\sum_k\bar{\rho}_{22}^0({\mit\Gamma};k) \,,
     \nonumber \\
   &&p_{1:n;2:n}
        = \frac{2}{L}\sum_k
          \cos\left(\frac{k}{2}\right)\,
          \bar{\rho}_{21}^0({\mit\Gamma};k)
         -\frac{2}{L}\sum_k
          \sin\left(\frac{k}{2}\right)\,
          \tilde{\rho}_{21}^0({\mit\Gamma};k) \,,
     \nonumber \\   
   &&p_{1:n;2:n-1}
        = \frac{2}{L}\sum_k
          \cos\left(\frac{k}{2}\right)\,
          \bar{\rho}_{21}^0({\mit\Gamma};k)
         +\frac{2}{L}\sum_k
          \sin\left(\frac{k}{2}\right)\,
          \tilde{\rho}_{21}^0({\mit\Gamma};k) \,,
     \nonumber \\   
   &&u_n= \frac{8\alpha}{L}\sum_k
          \sin\left(\frac{k}{2}\right)\,
          \tilde{\rho}_{21}^0 \,.
   \label{E:GB00}
\end{eqnarray}
This phase (Fig.$\!$ \ref{F:BSP}b) is characterized by the alternating
bond orders with the uniform lattice deformation.
We abbreviate the state as BOW.
Here is no alternation of on-site charge densities and the
site-diagonal el-ph coupling $\beta$ is irrelevant.

\subsection{$XA\otimes\check{S}^0\otimes\check{R}^0$}

\begin{eqnarray}
   &&d_{1:n}= \frac{2}{L}\sum_k
              \bar{\rho}_{11}^0({\mit\Gamma};k)
             +\frac{2(-1)^n}{L}\sum_k
              \bar{\rho}_{11}^0(X:k) \,,\ \
     d_{2:n}= \frac{2}{L}\sum_k
              \bar{\rho}_{22}^0({\mit\Gamma};k) \,,
     \nonumber \\
   &&p_{1:n;2:n}=p_{1:n;2:n-1}
        = \frac{2}{L}\sum_k
          \cos\left(\frac{k}{2}\right)\,
          \bar{\rho}_{21}^0({\mit\Gamma};k)
         -\frac{2(-1)^n}{L}\sum_k
          \sin\left(\frac{k}{2}\right)\,
          \bar{\rho}_{21}^0(X;k) \,,
      \nonumber \\
   &&u_n=-\frac{4\beta(-1)^n}{L}\sum_k
          \bar{\rho}_{11}^0(X:k)\,.
   \label{E:XA00}
\end{eqnarray}
This phase (Fig.$\!$ \ref{F:BSP}c) is characterized by the alternating
charge densities on the metal sites induced by the lattice
dimerization.
We abbreviate the state as M-CDW.
While both site-diagonal ($\beta$) and site-off-diagonal
($\alpha$) el-ph interactions contribute to the energy stabilization
of the state, $\beta$ is much more relevant here.

\subsection{$XB\otimes\check{S}^0\otimes\check{T}^0$}

\begin{eqnarray}
   &&d_{1:n}= \frac{2}{L}\sum_k\bar{\rho}_{11}^0({\mit\Gamma};k)\,,\ \
     d_{2:n}= \frac{2}{L}\sum_k\bar{\rho}_{22}^0
             +\frac{2(-1)^n}{L}\sum_k\bar{\rho}_{22}^0(X;k) \,,
     \nonumber \\
   &&p_{1:n;2:n}
        = \frac{2}{L}\sum_k
          \cos\left(\frac{k}{2}\right)\,
          \bar{\rho}_{21}^0({\mit\Gamma};k)
         -\frac{2(-1)^n}{L}\sum_k
          \sin\left(\frac{k}{2}\right)\,
          \bar{\rho}_{21}^1(X;k) \,,
     \nonumber \\
   &&p_{1:n;2:n-1}
        = \frac{2}{L}\sum_k
          \cos\left(\frac{k}{2}\right)\,
          \bar{\rho}_{21}^0({\mit\Gamma};k)
         +\frac{2(-1)^n}{L}\sum_k
          \sin\left(\frac{k}{2}\right)\,
          \bar{\rho}_{21}^1(X;k) \,.
   \label{E:XB00}
\end{eqnarray}
This phase (Fig.$\!$ \ref{F:BSP}d) is characterized by the alternating
charge densities on the halogen sites with a purely electronic bond
order wave.
We abbreviate the state as X-CDW.

\subsection{${\mit\Gamma}A\otimes\check{S}^1\otimes\check{T}^1$}

\begin{eqnarray}
   &&d_{1:n}=\frac{2}{L}\sum_k\bar{\rho}_{11}^0({\mit\Gamma};k)\,,\ \
     d_{2:n}=\frac{2}{L}\sum_k\bar{\rho}_{22}^0({\mit\Gamma};k)\,,
     \nonumber \\
   &&p_{1:n;2:n}=p_{1:n;2:n-1}
        =\frac{2}{L}\sum_k
         \cos\left(\frac{k}{2}\right)\,
         \bar{\rho}_{21}^0({\mit\Gamma};k)\,,
     \nonumber \\
   &&s_{1:n}^z=\frac{1}{L}\sum_k\bar{\rho}_{11}^z({\mit\Gamma};k)\,,\ \
     s_{2:n}^z=\frac{1}{L}\sum_k\bar{\rho}_{22}^z({\mit\Gamma};k)\,,
     \nonumber \\
   &&t_{1:n;2:n}^z=t_{1:n;2:n-1}^z
        =\frac{1}{L}\sum_k
         \cos\left(\frac{k}{2}\right)\,
         \bar{\rho}_{21}^z({\mit\Gamma};k) \,.
   \label{GA11}
\end{eqnarray}
This phase (Fig.$\!$ \ref{F:BSP}e) is the ferromagnetism with the
uniform spin bond orders.
We abbreviate the state as FM.
We note that not only the metal sites but also the halogen sites have
spin densities.

\subsection{${\mit\Gamma}B\otimes\check{S}^1\otimes\check{T}^1$}

\begin{eqnarray}
   &&d_{1:n}=\frac{2}{L}\sum_k\bar{\rho}_{11}^0({\mit\Gamma};k)\,,\ \
     d_{2:n}=\frac{2}{L}\sum_k\bar{\rho}_{22}^0({\mit\Gamma};k)\,,
     \nonumber \\
   &&p_{1:n;2:n}=p_{1:n;2:n-1}
        =\frac{2}{L}\sum_k
         \cos\left(\frac{k}{2}\right)\,
         \bar{\rho}_{21}^0({\mit\Gamma};k)\,,
     \nonumber \\
   &&t_{1:n;2:n}^z=-t_{1:n;2:n-1}^z
        =-\frac{1}{L}\sum_k
         \sin\left(\frac{k}{2}\right)\,
         \tilde{\rho}_{21}^z({\mit\Gamma};k) \,.
   \label{GB11}
\end{eqnarray}
This phase (Fig.$\!$ \ref{F:BSP}f) is characterized by the alternating
spin bond orders.
Both metal and halogen sites have no spin density.
We abbreviate the state as SBOW.

\subsection{$XA\otimes\check{S}^1\otimes\check{T}^1$}

\begin{eqnarray}
   &&d_{1:n}=\frac{2}{L}\sum_k\bar{\rho}_{11}^0({\mit\Gamma};k)\,,\ \
     d_{2:n}=\frac{2}{L}\sum_k\bar{\rho}_{22}^0({\mit\Gamma};k)\,,
     \nonumber \\
   &&p_{1:n;2:n}=p_{1:n;2:n-1}
        =\frac{2}{L}\sum_k
         \cos\left(\frac{k}{2}\right)\,
         \bar{\rho}_{21}^0({\mit\Gamma};k)\,,
     \nonumber \\
   &&s_{1:n}^z=\frac{(-1)^n}{L}\sum_k\bar{\rho}_{11}^z(X;k)\,,\ \
     s_{2:n}^z=0 \,,
     \nonumber \\
   &&t_{1:n;2:n}^z=t_{1:n;2:n-1}^z
        =\frac{(-1)^n}{L}\sum_k
         \cos\left(\frac{k}{2}\right)\,
         \bar{\rho}_{21}^z(X;k) \,.
   \label{XA11}
\end{eqnarray}
This phase (Fig.$\!$ \ref{F:BSP}g) is characterized by the alternating
spin densities on the metal sites with the alternating spin bond
orders.
We abbreviate the state as M-SDW.

\subsection{$XB\otimes\check{S}^1\otimes\check{T}^1$}

\begin{eqnarray}
   &&d_{1:n}=\frac{2}{L}\sum_k\bar{\rho}_{11}^0({\mit\Gamma};k)\,,\ \
     d_{2:n}=\frac{2}{L}\sum_k\bar{\rho}_{22}^0({\mit\Gamma};k)\,,
     \nonumber \\
   &&p_{1:n;2:n}=p_{1:n;2:n-1}
        =\frac{2}{L}\sum_k
         \cos\left(\frac{k}{2}\right)\,
         \bar{\rho}_{21}^0({\mit\Gamma};k)\,,
     \nonumber \\
   &&s_{1:n}^z=0\,,\ \
     s_{2:n}^z=\frac{(-1)^n}{L}\sum_k\tilde{\rho}_{22}^z(X;k)\,,
     \nonumber \\
   &&t_{1:n;2:n}^z=-t_{1:n;2:n-1}^z
        =-\frac{(-1)^n}{L}\sum_k
          \sin\left(\frac{k}{2}\right)\,
          \tilde{\rho}_{21}^z(X;k) \,.
   \label{XB11}
\end{eqnarray}
This phase (Fig.$\!$ \ref{F:BSP}h) is characterized by the alternating
spin densities on the halogen sites with the alternating spin bond
orders.
We abbreviate the state as X-SDW.
\narrowtext

   So-far-synthesized MX compounds exhibit ground states of M-CDW
or M-SDW type.
We will later visualize the competition \cite{Nasu98} between them,
which is essentially described by the two parameters, $\beta$ and
$U_{\rm M}$.
The ferroelectric ground states \cite{Torr53,Toku05} of mixed-stack
CT compounds are recognized as BOW, where $\alpha$ is most relevant.
The ferromagnetic ground states \cite{Kino25} of organic radical
crystals may qualitatively be identified with FM, where strong
on-site Coulomb interactions are essential.
Thus it is implied that M-CDW and M-SDW are most stabilized at $3/4$
band filling, whereas BOW and FM around half band filling.
Generally, $\mit\Gamma$- and $X$-phases are, respectively, more and
less stabilized as the system moves away from $3/4$ band filling.
Our approach shows that only the nonmagnetic solutions can be
accompanied by lattice deformation.
Further harvests may be a few novel density-wave states, X-CDW,
SBOW, and X-SDW.
We note that M-CDW is stabilized in a lattice dimerized, whereas no
lattice distortion accompanies X-CDW.
It is also worth while noting that the three-band model for the
CuO$_2$ planes in high-temperature superconductors also has a
solution of the same type as X-SDW, namely, the SDW on the oxygen
sites \cite{Yama29,Yama99,Wein35}.
Assuming that doped holes should predominantly be of halogen
character on the analogy of the model in two dimension \cite{Dutk05},
hole doping may cause the frustration of an antiferromagnetic spin
alignment on the metal sites and therefore lead to the collapse of
M-SDW.

\section{Numerical Calculation and Discussion}\label{S:NCD}

   Plenty of parameters may potentially bring rich ground-state
phase diagrams.
In order to have a general view of the model, we here restrict
relevant parameters to the difference between the M- and X-atom
on-site affinities,
$\varepsilon_0\equiv\varepsilon_{\rm M}-\varepsilon_{\rm X}$,
the on-site Coulomb interactions, $U_{\rm M}$ and $U_{\rm X}$, and
band filling.
The HF energy is calculated in the thermodynamic limit.

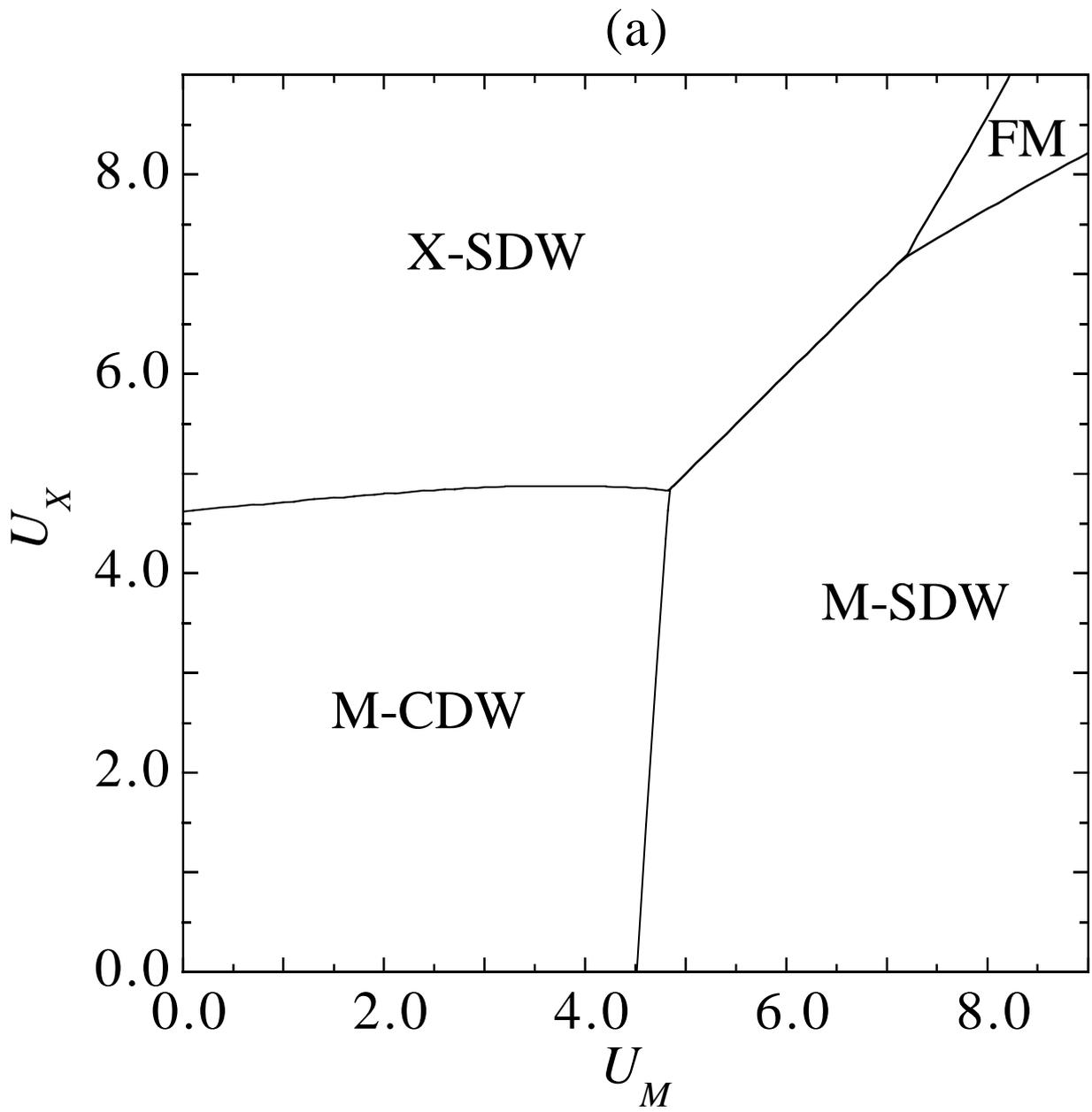
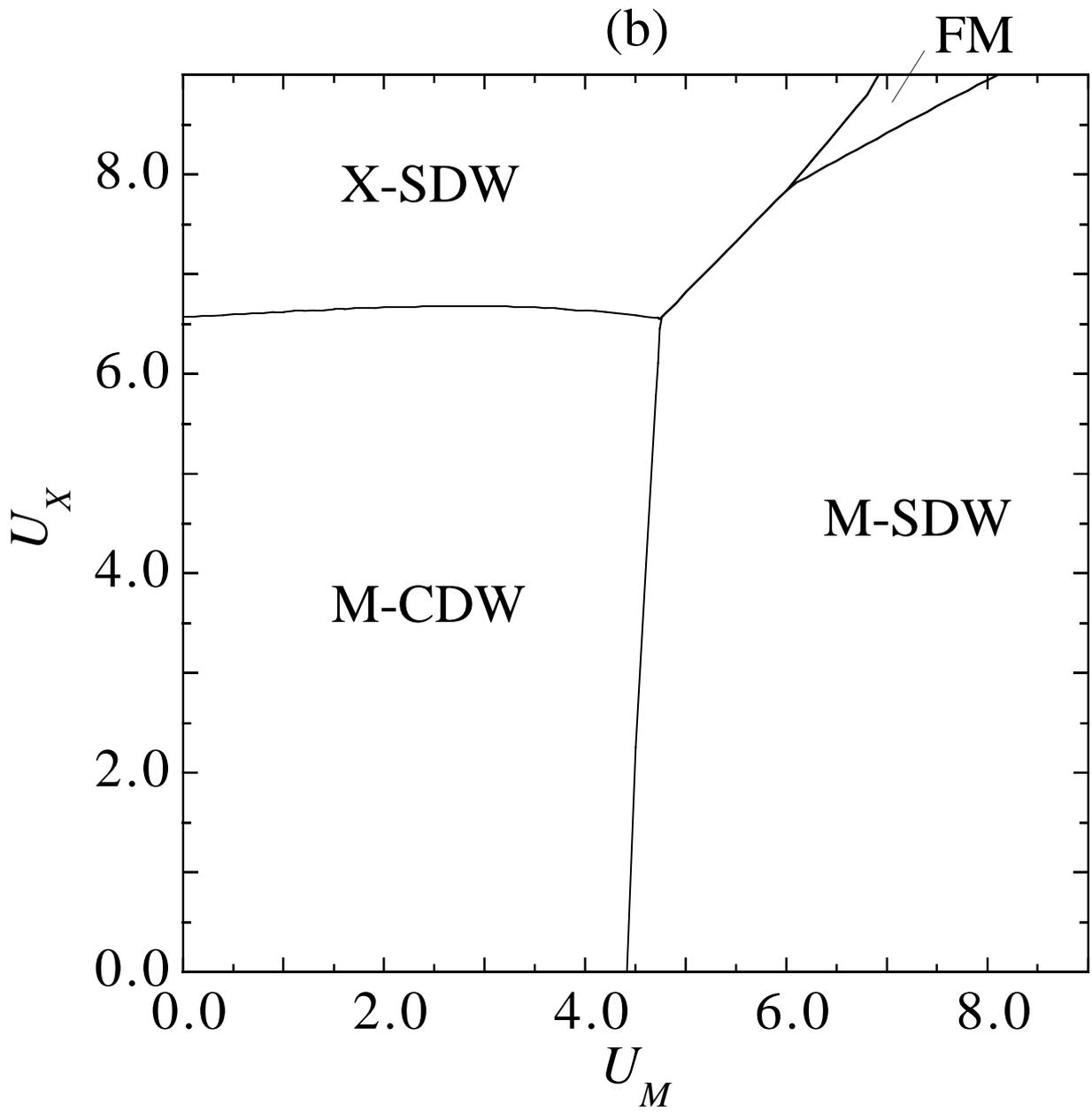
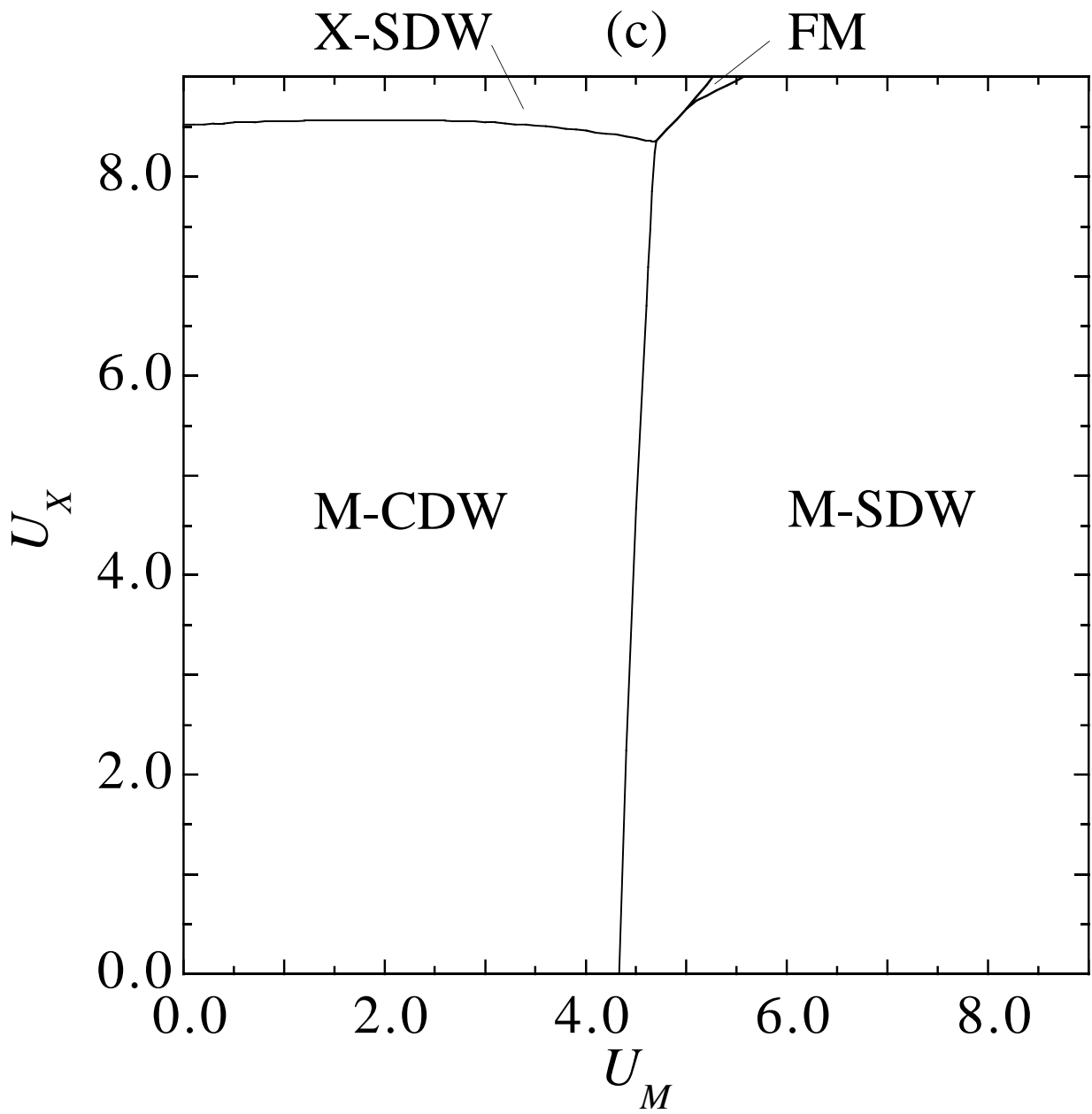
\begin{figure}
\caption{Ground-state phase diagrams at various values of
         $\varepsilon_0$:
         (a) $\varepsilon_0/t_0=0.0$;
         (b) $\varepsilon_0/t_0=1.0$;
         (c) $\varepsilon_0/t_0=2.0$.
         Here the following parametrization is common to all:
         $V/t_0=1.0$,
         $\alpha/(Kt_0)^{1/2}=0.1$,
         $\beta/(Kt_0)^{1/2}=1.0$, and
         $3/4$ band filling.}
\label{F:PhD1}
\end{figure}

   We show in Fig. \ref{F:PhD1} phase diagrams at $3/4$ band
filling, where $q=\pi$ states are predominantly stabilized.
As far as magnetic instabilities are concerned,
the phase diagram at $\varepsilon_0=0$ is almost symmetric under the
exchange of $U_{\rm M}$ and $U_{\rm X}$.
This is because magnetic phases are accompanied by no lattice
deformation.
It is not the case with nonmagnetic instabilities.
Here the metal-atom displacements are assumed to be much smaller
than the halogen-atom ones and thus negligible.
Therefore, X-CDW is strongly suppressed and M-CDW is predominantly
stabilized due to the el-ph coupling $\beta$.
With the increase of $\varepsilon_0$, density waves on the halogen
sites are generally reduced and single-band models come to be
justified.
The phase boundary between M-SDW and X-SDW is roughly given by
$U_{\rm X}-U_{\rm M}=1.8\varepsilon_0$.
In comparison with MX compounds with ${\rm M}={\rm Pt}$ and
${\rm M}={\rm Pd}$ whose ground states are M-CDW, Ni complexes
relatively have a small $\varepsilon_0$ and a large $U_{\rm M}$
\cite{Okam38} and thus exhibit M-SDW ground states.
In order to realize X-SDW ground states, a small enough
$\varepsilon_0$ and a large enough $U_{\rm X}$ are necessary.

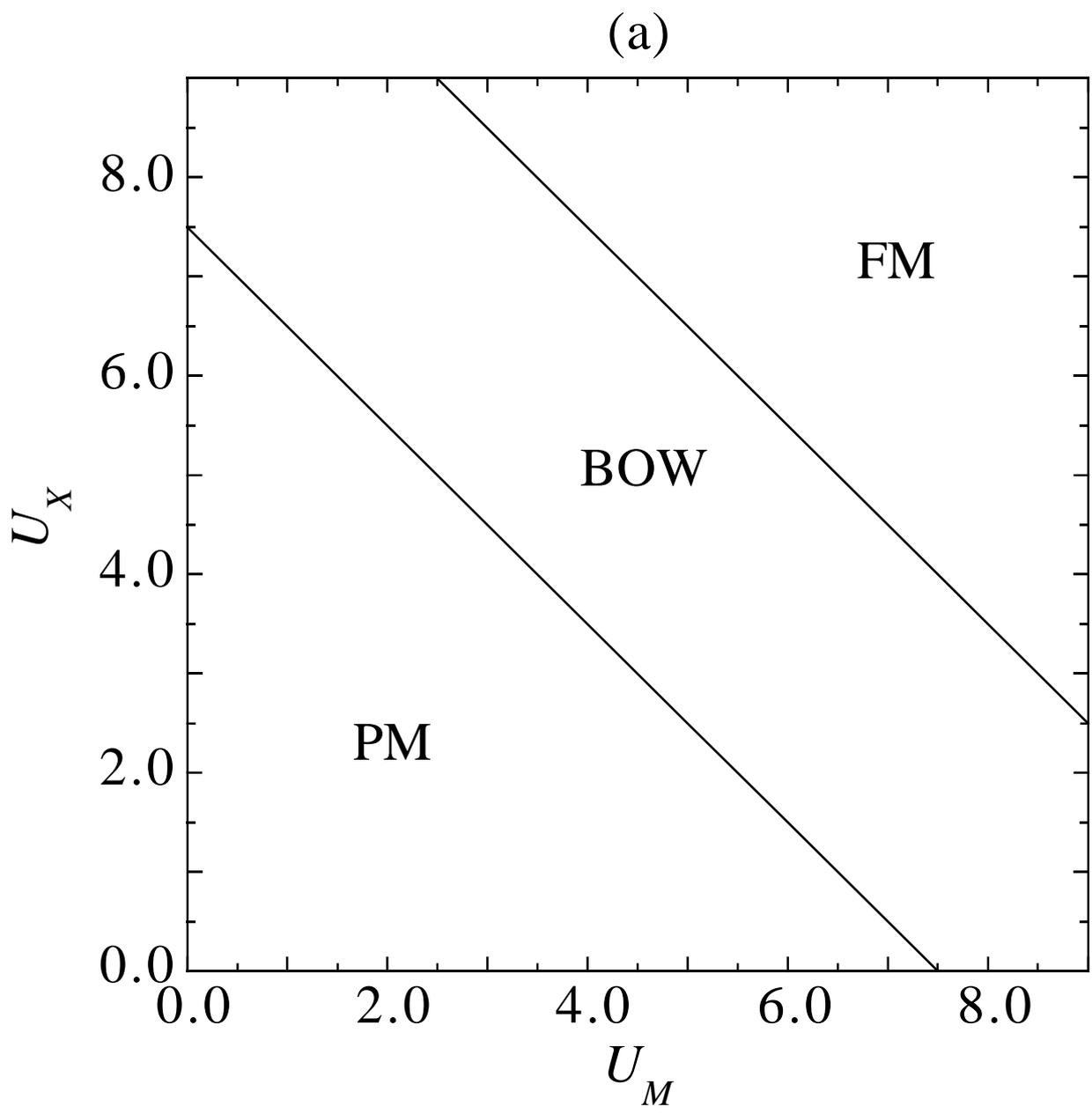
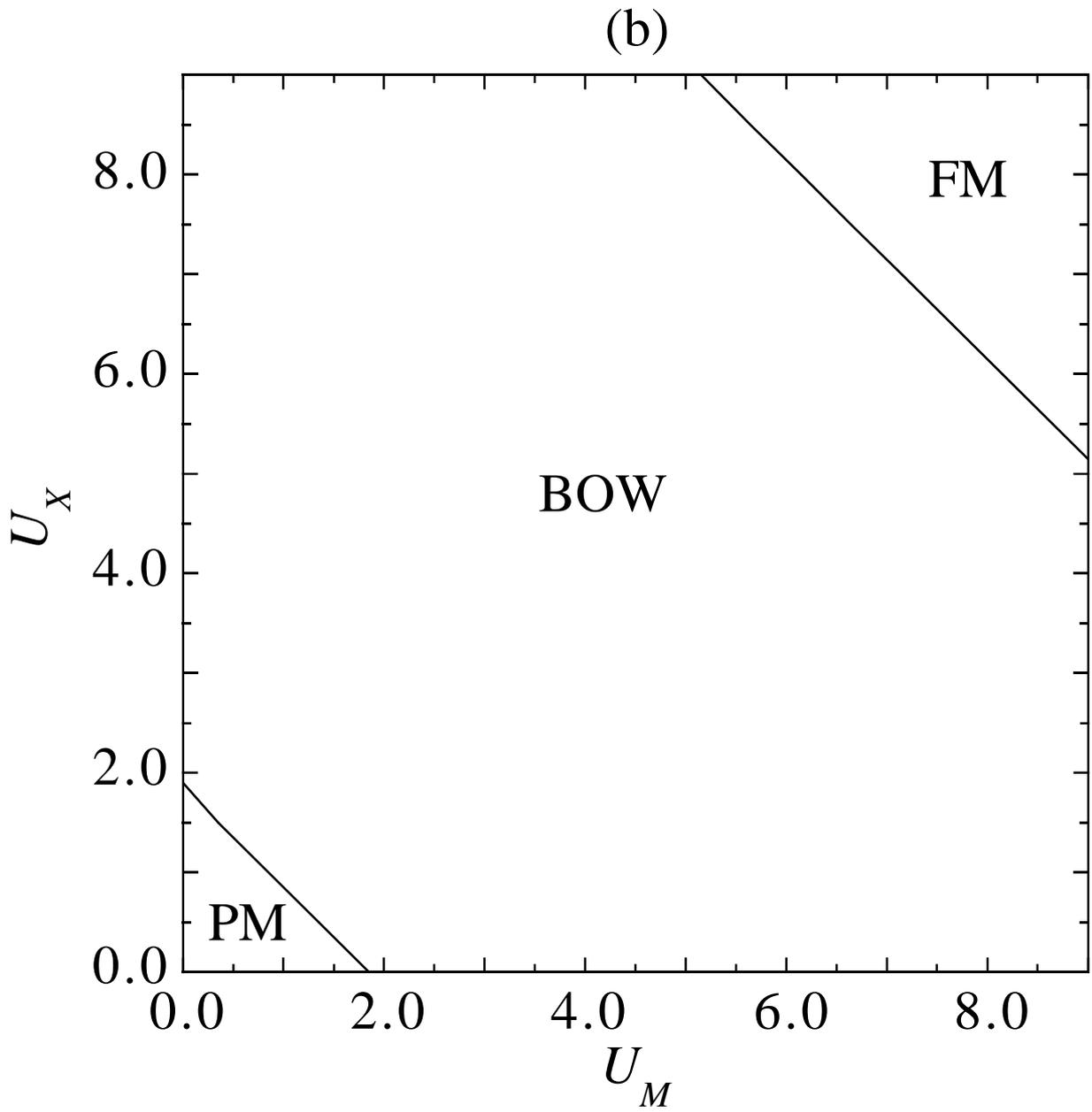
\begin{figure}
\caption{Ground-state phase diagrams at various values of
         el-ph coupling constants:
         (a) $\alpha/(Kt_0)^{1/2}=0.6$, $\beta/(Kt_0)^{1/2}=1.0$;
         (b) $\alpha/(Kt_0)^{1/2}=0.8$, $\beta/(Kt_0)^{1/2}=0.1$.
         Here the following parametrization is common to all:
         $\varepsilon_0/t_0=3.0$,
         $V/t_0=1.0$, and
         half band filling.}
\label{F:PhD2}
\end{figure}

   We show in Fig. \ref{F:PhD2} phase diagrams at half band
filling, where $q=0$ states are relatively stabilized.
The off-site el-ph coupling $\alpha$ stabilizes BOW, whereas
the off-site Coulomb interaction $V$ is unfavorable to that.
The increase of $V$ generally turns BOW into PM.
All the present findings are consistent with the observations
\cite{Torr53,Toku05} of mixed-stack CT compounds.
Here we could not have a quantitative argument on FM.
The present approach never goes beyond the Stoner's theory
\cite{Ston43} and thus concludes that once the system deviates
from $3/4$ band filling, FM can necessarily appear with large enough
$U_{\rm M}$ and $U_{\rm X}$.
It is, however, well-known that multiscattering effects \cite{Kana75}
qualitatively correct this scenario.
The ferromagnetism in multiband models \cite{Shib71,Naga16,Kusa17}
is a subject of great interest in itself and should be
investigated taking account of intra-atomic exchange interactions,
which favors a ferromagnetic spin alignment.

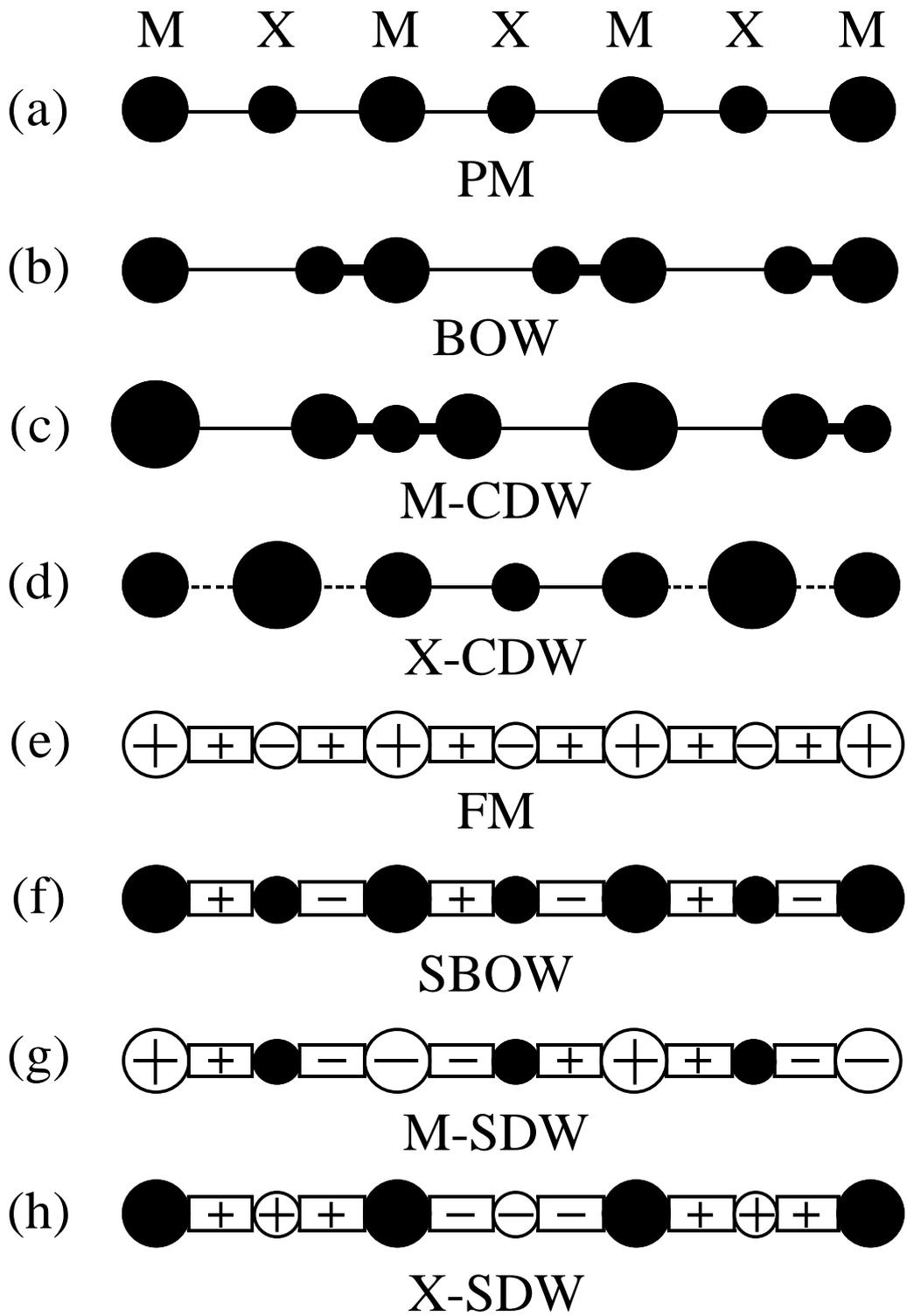
\begin{figure}
\caption{Schematic representation of possible density-wave states,
         where the variety of circles and segments qualitatively
         represents the variation of local charge densities and
         bond orders, respectively, whereas the signs $\pm$ in
         circles and strips describe the alternation of local spin
         densities and spin bond orders, respectively.
         Circles shifted from their equidistant location
         qualitatively represent X-atom displacements.
         Identifying the present system, for example, with
         halogen-bridged metal complexes, each phase is characterized
         as follows:
         (a) Paramagnetism;
         (b) Electron-phonon bond order wave;
         (c) Metal charge density wave accompanied by an
             electron-phonon bond order wave;
         (d) Halogen charge density wave accompanied by a purely
             electronic bond order wave;
         (e) Ferromagnetism with uniform spin bond orders;
         (f) Spin bond order wave;
         (g) Metal spin density wave accompanied by a
             spin bond order wave;
         (h) Halogen spin density wave accompanied by a
             spin bond order wave.}
\label{F:BSP}
\end{figure}

   Finally we briefly observe the doping dependences of the states
appearing in the phase diagrams.
We define macroscopic order parameters as
\begin{eqnarray}
   &&
   O_{\rm M\mbox{-}CDW}
     =\frac{1}{L}
      \sum_n
      (-1)^n d_{1:n}\,,
   \label{E:OPMCDW} \\
   &&
   O_{\rm M\mbox{-}SDW}
     =\frac{1}{L}
      \sum_n
      (-1)^n s_{1:n}^z\,,
   \label{E:OPMSDW} \\
   &&
   O_{\rm X\mbox{-}SDW}
     =\frac{1}{L}
      \sum_n
      (-1)^n s_{2:n}^z\,,
   \label{E:OPXSDW} \\
   &&
   O_{\rm BOW}
     =\frac{1}{2L}
      \sum_n
      \left(
       p_{1:n;2:n-1}-p_{1:n;2:n}
      \right)\,,
   \label{E:OPBOW} \\
   &&
   O_{\rm FM}^{\rm M}
     =\frac{1}{L}
      \sum_n
      s_{1:n}^z\,,\ \ 
   O_{\rm FM}^{\rm X}
     =\frac{1}{L}
      \sum_n
      s_{2:n}^z\,,
   \label{E:OPFM}
\end{eqnarray}
and show them as functions of band filling in Table \ref{T:OP},
where the last digit of each estimate may contain a slight numerical
uncertainty.
Now we explicitly find that M-CDW, M-SDW, and X-SDW are most
stabilized at $3/4$ band filling, while BOW and FM at half band
filling.
Concerning FM we note that $O_{\rm FM}^{\rm M}$ and
$O_{\rm FM}^{\rm X}$ have opposite signs, which means spins on the
metal and halogen sites are antiparallel to each other
(Fig.$\!$ \ref{F:BSP}e).
When we move away from half band filling, electrons are preferably
doped into the halogen sites, whereas holes into the metal sites.
Further numerical investigation will be presented elsewhere.

   While X-CDW and SBOW do not appear in the ground-state phase
diagrams shown here, we expect them in a more extensive numerical
investigation.
Even if the states are scarcely stabilized into a ground state, they
can still be relevant in the ground-state correlations.
We note, for example, that low-lying solitonic excitations in a SDW
ground state induce SBOW domains around their centers \cite{Shim07}.
Besides the obtained novel density-wave states themselves, we stress
the wide applicability of the present approach.
Although the approach is feasible and even more enlightening in
higher dimensions \cite{Yama29,Yama99}, fascinating materials such
as halogen-bridged bi-nuclear complexes \cite{Mita00} and DCNQI-Cu
systems exhibiting a three-fold superlattice structure \cite{Kato24}
stimulate us to further explorations in one dimension, where the
problem of competing ground states is really fruitful.
Even if a model treated is of one dimension, its complicated unit-cell
structure and various el-ph interactions may conceal some of
instabilities of importance from our naive guess.
Our approach never fails to reveal all the
possible broken-symmetry phases.
We hope that the present argument will motivate further chemical,
as well as theoretical, explorations in low-dimensional el-ph
systems.

\acknowledgments

   It is a pleasure to thank M. Ozaki for his useful discussion.
The author is furthermore grateful to Y. Nogami and H. Okamoto for
their helpful comments on materials of current interest. 
A part of the numerical calculation was done using the facility of the
supercomputer center, Institute for Solid State Physics, University of
Tokyo.
This work is supported in part by the Japanese Ministry of Education,
Science, and Culture through the Grant-in-Aid 09740286 and by the
Okayama Foundation for Science and Technology.

\widetext
\newpage

\begin{table}
\caption{Hartree-Fock broken-symmetry Hamiltonians, their invariance
         groups, and consequent density-wave states.}
\begin{tabular}{clll}
Representation & Hamiltonian & Invariance group & State \\
\tableline
\noalign{\vskip 2pt}
${\mit\Gamma}A\otimes\check{S}^0\otimes\check{T}^0$ &
$h^0({\mit\Gamma};A)$                               &
${\bf L}_1{\bf C}_2{\bf S}{\bf T}$                  &
PM                                                  \\
${\mit\Gamma}B\otimes\check{S}^0\otimes\check{T}^0$ &
$h^0({\mit\Gamma};A)+h^0({\mit\Gamma};B)$           &
${\bf L}_1{\bf S}{\bf T}$                           &
BOW                                                 \\
$X           A\otimes\check{S}^0\otimes\check{T}^0$ &
$h^0({\mit\Gamma};A)+h^0(X           ;A)$           &
${\bf L}_2{\bf C}_2{\bf S}{\bf T}$                  &
M-CDW                                               \\
$X           B\otimes\check{S}^0\otimes\check{T}^0$ &
$h^0({\mit\Gamma};A)+h^0(X           ;B)$           &
$(1+l_1C_2){\bf L}_2{\bf S}{\bf T}$                 &
X-CDW                                               \\
${\mit\Gamma}A\otimes\check{S}^1\otimes\check{T}^1$ &
$h^0({\mit\Gamma};A)+h^z({\mit\Gamma};A)$           &
${\bf L}_1{\bf C}_2
 {\bf A}(\mbox{\boldmath$e$}_z)
 {\bf M}(\mbox{\boldmath$e$}_\parallel)$            &
FM                                                  \\
${\mit\Gamma}B\otimes\check{S}^1\otimes\check{T}^1$ &
$h^0({\mit\Gamma};A)+h^z({\mit\Gamma};B)$           &
$(1+C_2u(\mbox{\boldmath$e$}_\parallel,\pi))
 {\bf L}_1
 {\bf A}(\mbox{\boldmath$e$}_z)
 {\bf M}(\mbox{\boldmath$e$}_\parallel)$            &
SBOW                                                \\
$X           A\otimes\check{S}^1\otimes\check{T}^1$ &
$h^0({\mit\Gamma};A)+h^z(X           ;A)$           &
$(1+l_1u(\mbox{\boldmath$e$}_\parallel,\pi))
 {\bf L}_2{\bf C}_2
 {\bf A}(\mbox{\boldmath$e$}_z)
 {\bf M}(\mbox{\boldmath$e$}_\parallel)$            &
M-SDW                                               \\
$X           B\otimes\check{S}^1\otimes\check{T}^1$ &
$h^0({\mit\Gamma};A)+h^z(X           ;B)$           &
$(1+l_1C_2)
 (1+l_1u(\mbox{\boldmath$e$}_\parallel,\pi))
 {\bf L}_2
 {\bf A}(\mbox{\boldmath$e$}_z)
 {\bf M}(\mbox{\boldmath$e$}_\parallel)$            &
X-SDW                                               \\
\end{tabular}
$^{\rm a)}$${\bf L}_1$ is the translation group whose basis
vector is the unit-cell translation $l_1$.
\hfill\break
$^{\rm b)}$${\bf L}_2$ is the translation group whose basis
vector is the double-unit-cell translation.
\hfill\break
$^{\rm c)}$${\bf A}(\mbox{\boldmath$e$}_z)
            =\{u(\mbox{\boldmath$e$}_z,\theta)
            \vert 0\leq\theta\leq 4\pi\}$
with the axis $\mbox{\boldmath$e$}_z$ along the $z$ direction.
\hfill\break
$^{\rm d)}$${\bf M}(\mbox{\boldmath$e$}_\parallel)
            =\{E,tu(\mbox{\boldmath$e$}_\parallel,\pi)\}$
with the axis $\mbox{\boldmath$e$}_\parallel$ along the chain
direction.
\label{T:H}
\end{table}

\begin{table}
\caption{Order parameters as functions of band filling.}
\begin{tabular}{cccc}
\ Band filling\ & $O_{\rm M\mbox{-}CDW}$
                & $O_{\rm M\mbox{-}SDW}$
                & $O_{\rm X\mbox{-}SDW}$ \\
\noalign{\vskip 1pt}
\tableline
$3.2/4.0$ & $0.64977$ & $0.32626$ & $0.23942$ \\
$3.1/4.0$ & $0.74603$ & $0.37371$ & $0.28664$ \\
$3.0/4.0$ & $0.84207$ & $0.42111$ & $0.33321$ \\
$2.9/4.0$ & $0.78778$ & $0.40047$ & $0.32808$ \\
$2.8/4.0$ & $0.72453$ & $0.37621$ & $0.32137$ \\
\tableline
\noalign{\vskip 2pt}
\ Band filling\ & $O_{\rm BOW}$
                & $O_{\rm FM}^{\rm M}$
                & $O_{\rm FM}^{\rm X}$ \\
\noalign{\vskip 1pt}
\tableline
$2.2/4.0$ & $0.32294$ & $0.42921$ & $-0.32921$\ \ \\
$2.1/4.0$ & $0.46682$ & $0.43536$ & $-0.38536$\ \ \\
$2.0/4.0$ & $0.47283$ & $0.44118$ & $-0.44118$\ \ \\
$1.9/4.0$ & $0.47147$ & $0.38536$ & $-0.43536$\ \ \\
$1.8/4.0$ & $0.46954$ & $0.32921$ & $-0.42921$\ \
\end{tabular}
\label{T:OP}
\end{table}

\end{document}